\documentclass[sigconf]{acmart}
\usepackage{balance}
\usepackage{booktabs} 
\usepackage{amsmath,color}
\usepackage{makecell}
\usepackage{caption}
\usepackage{subcaption}
\usepackage{tablefootnote}
\usepackage{float}
\usepackage{tabularx}
\newcolumntype{b}{X} 
\newcolumntype{s}{>{\hsize=.5\hsize}X} 
\usepackage{flushend}
\usepackage{booktabs} 
\setcopyright{acmcopyright}

\newcommand{\squishlist}{
  \begin{list}{$\bullet$}
    { \setlength{\itemsep}{0pt}      \setlength{\parsep}{3pt}
      \setlength{\topsep}{3pt}       \setlength{\partopsep}{0pt}
      \setlength{\leftmargin}{1.5em} \setlength{\labelwidth}{1em}
      \setlength{\labelsep}{0.5em} } }
\newcommand{\squishlisttwo}{
  \begin{list}{$\bullet$}
    { \setlength{\itemsep}{0pt}    \setlength{\parsep}{0pt}
      \setlength{\topsep}{0pt}     \setlength{\partopsep}{0pt}
      \setlength{\leftmargin}{0.9em} \setlength{\labelwidth}{0.5em}
      \setlength{\labelsep}{0.5em} } }

 \newcommand{\squishend}{
     \end{list} 
 }
 
\newenvironment{rcases}
{\left.\begin{aligned}}
{\end{aligned}\right\rbrace}

\copyrightyear{2018} 
\acmYear{2018} 
\setcopyright{acmcopyright}
\acmConference[WSDM 2018]{WSDM 2018: The Eleventh ACM International Conference on Web Search and Data Mining }{February 5--9, 2018}{Marina Del Rey, CA, USA}
\acmPrice{15.00}
\acmDOI{10.1145/3159652.3159693}
\acmISBN{978-1-4503-5581-0/18/02}
\hypersetup{draft}

\begin{document}
\title{VISIR: Visual and Semantic Image Label Refinement}

\author{Sreyasi Nag Chowdhury\texorpdfstring{\textsuperscript{*}}{}\hspace{0.75cm} Niket Tandon\texorpdfstring{\textsuperscript{$\dagger$}}{}\hspace{0.75cm} Hakan Ferhatosmanoglu\texorpdfstring{\textsuperscript{$\ddagger$}}{}\hspace{0.75cm} Gerhard Weikum\texorpdfstring{\textsuperscript{*}\\}{}
\mbox{\small \hspace{0.6cm} sreyasi@mpi-inf.mpg.de\hspace{1.7cm}nikett@allenai.org\hspace{1.9cm}hakan.f@warwick.ac.uk\hspace{1.8cm}weikum@mpi-inf.mpg.de}}
\affiliation{%
	\institution{\textsuperscript{*}Max Planck Institute for Informatics\hspace{0.7cm}
							 \textsuperscript{$\dagger$} Allen Institute for Artificial Intelligence\hspace{0.7cm}
							 \textsuperscript{$\ddagger$} University of Warwick}
}

\begin{abstract}
The social media explosion has populated the Internet with a wealth of images. 
There are two existing paradigms for image retrieval:
1)~content-based image retrieval (CBIR), which has traditionally used visual features for similarity search (e.g., SIFT features), and 
2)~tag-based image retrieval (TBIR), which has relied on user tagging (e.g., Flickr tags). 
CBIR now gains semantic expressiveness by advances
in deep-learning-based detection of visual labels.
TBIR benefits from query-and-click logs to
automatically infer more informative labels. 
However, learning-based tagging still yields
noisy labels and
is restricted to concrete objects,
missing out
on generalizations and abstractions.
Click-based tagging is limited to terms that
appear in the textual context of an image or in queries that lead to a click.
This paper addresses the above limitations
by semantically refining and expanding the labels 
suggested by learning-based object detection.
We consider the semantic coherence between the labels for
different objects, leverage lexical and commonsense knowledge,
and cast the label assignment into a constrained optimization problem
solved by an integer linear program.
Experiments show that our method,
called VISIR, improves the quality of the state-of-the-art visual labeling tools like LSDA and YOLO.
\end{abstract}




\maketitle

\section{Introduction}
\noindent
\textbf{Motivation and Problem:} 
The enormous growth of social media has populated the Internet with a wealth of images.
On one hand, this makes image search easier, as there is redundancy for
many keywords with informative text surrounding the images.
On the other hand, it makes search harder, as there is a huge
amount of visual contents that is hardly understood by the search engine.
There are two paradigms for searching images:
content-based image retrieval (CBIR) and tag-based image retrieval (TBIR).

\begin{figure}[t]
    \centering
    \begin{subfigure}[t]{0.5\columnwidth}
        \includegraphics[width=4cm]{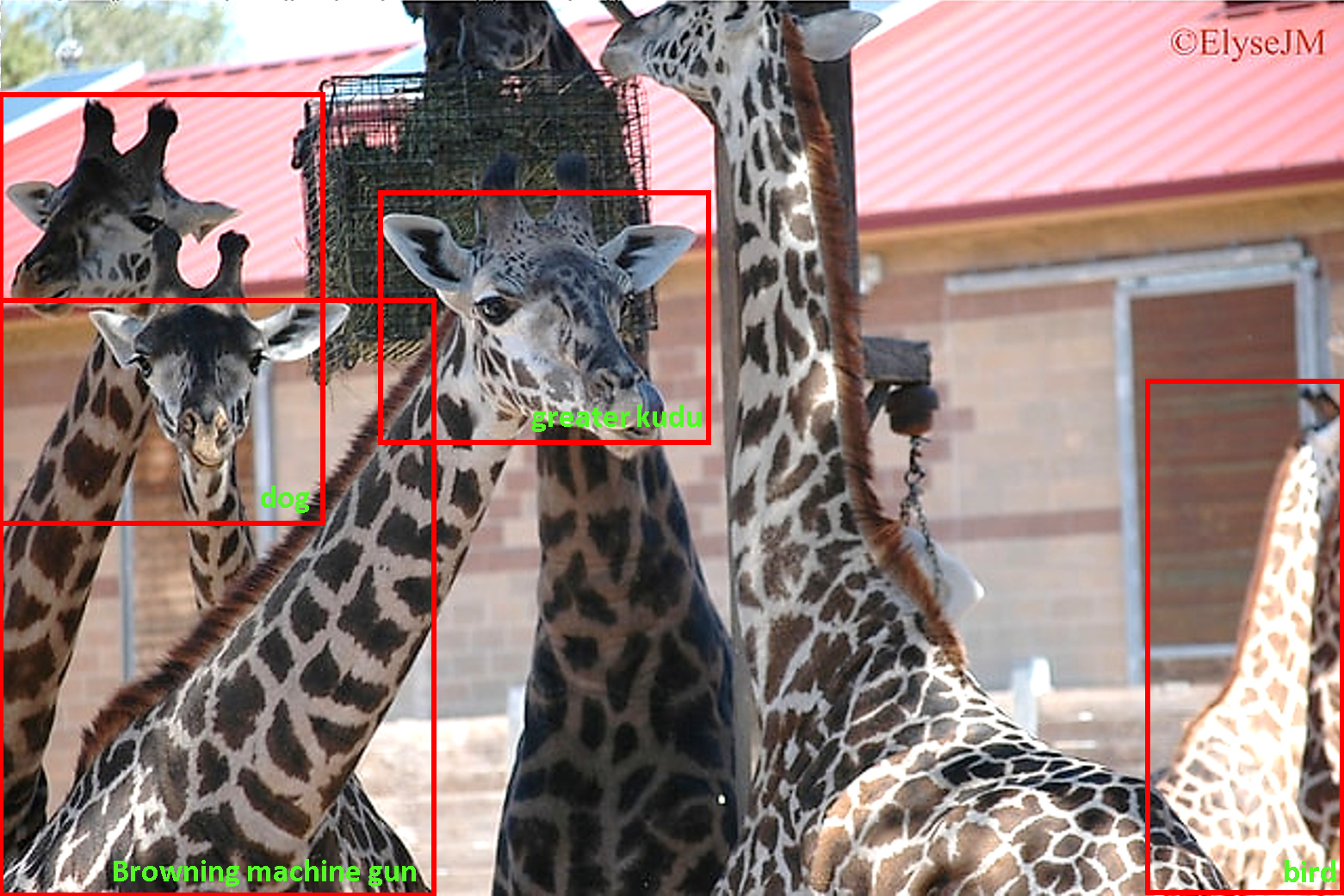}
        \caption{}\label{fig:cbir}
    \end{subfigure}%
    ~
    \begin{subfigure}[t]{0.5\columnwidth}
        \flushright
        \includegraphics[width=4cm]{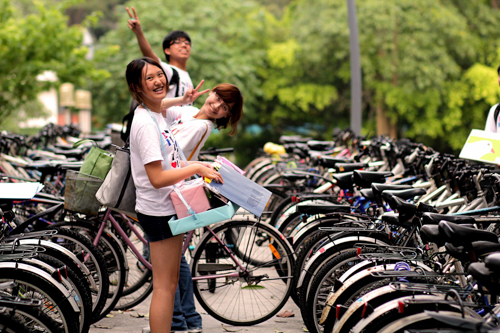}
        \caption{}\label{fig:tbir}
    \end{subfigure}
 \vspace*{-0.5cm}
    \caption{\small Noisy and Incomplete Labels:
    a) from LSDA~\cite{hoffman2014lsda} - {\em dog, Browning machine gun, greater kudu, bird} \hspace{0.2em}
    b) from flickr.com - {\em happiness}}
    \label{existingIssues}
    \vspace{-0.5cm}
\end{figure}

\begin{figure}[!b]
    \begin{tabular}{|p{2.6cm}|p{2.4cm}|p{2.4cm}|}
    \hline
            &  \vspace{0.05em}\hspace{1em}LSDA Labels & \vspace{0.05em}\hspace{1em}VISIR labels\\
    \hline
         \raisebox{-\totalheight}{\includegraphics[width=8.3em]{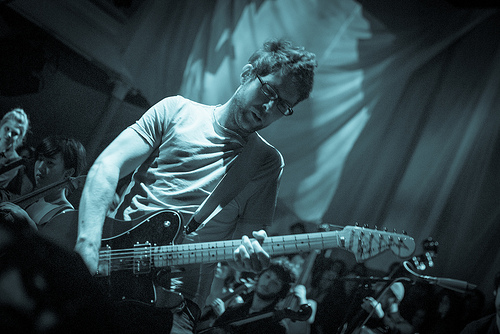}}\vspace{0.5em} & 
         \vspace{0.05em}\makecell{{\color{ACMRed} allosaurus}\\{\color{ACMRed}loggerhead turtle}\\person\\ {\color{ACMRed} bird}} & 
         \makecell{person\\ guitar\\ {\color{ACMBlue}\textit{stringed instrument}} \\{\color{ACMGreen} \textit{self-expression}}} \\
    \hline
        \raisebox{-\totalheight}{\includegraphics[width=8.3em]{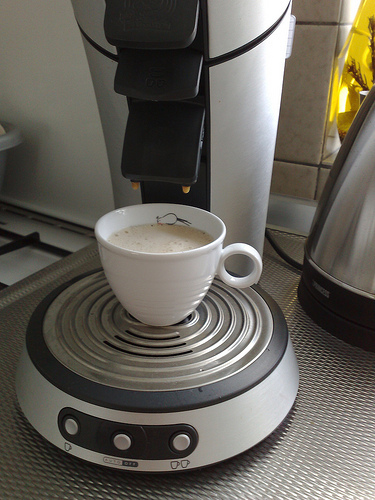}}\vspace{0.5em} & \vspace{0.1em}\makecell{bone china\\{\color{ACMRed}stove}\\{\color{ACMRed}WC, loo}\\cup or mug} & 
        \vspace{0.1em}\makecell{{\color{ACMRed} food processor}\\{\color{ACMRed} bowl}\\cup or mug\\{\color{ACMBlue}\textit{utensil}}}\\
    \hline
        \raisebox{-\totalheight}{\includegraphics[width=8.3em]{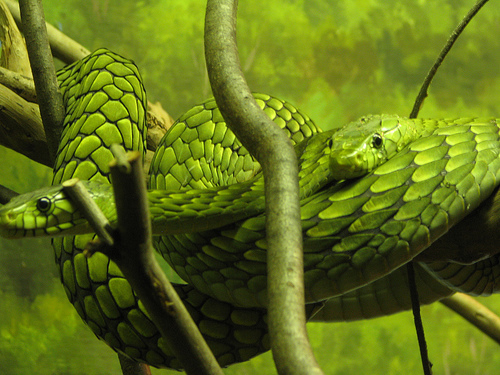}}\vspace{0.5em} & \vspace{0.1em}\makecell{{\color{ACMRed} cucumber}\\snake\\green mamba} & 
        \vspace{0.1em}\makecell{snake\\{\color{ACMBlue} \textit{reptile}}\\{\color{ACMGreen}\textit{slithery}}\\{\color{ACMGreen}\textit{poisonous}}} \\
    \hline
        \raisebox{-\totalheight}{\includegraphics[width=8.3em]{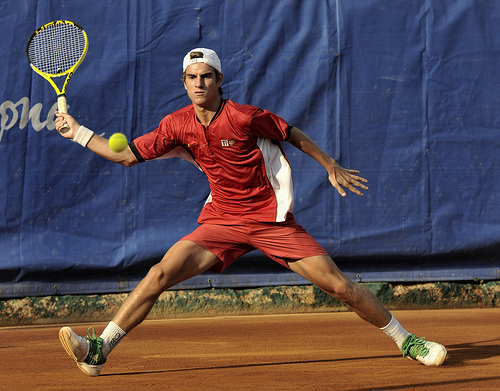}}\vspace{0.5em} & \vspace{0.1em}\makecell{racket\\person\\{\color{ACMRed}bathing cap}\\tennis ball\\{\color{ACMRed}head cabbage}} & 
        \vspace{0.1em}\makecell{tennis bat\\{\color{ACMBlue} \textit{individual}}\\{\color{ACMGreen}\textit{play tennis}}\\tennis ball} \\
    \hline
    \end{tabular}
    \caption{\small Images with labels from LSDA and VISIR}
    \label{goodExamples}
\end{figure}

CBIR finds images similar to a query image based on 
visual features that are used to represent an image. These features include color, shape, texture, SIFT descriptors etc. 
(e.g., 
\cite{DBLP:journals/tomccap/LewSDJ06,DBLP:journals/csur/DattaJLW08,DBLP:conf/cvpr/ArandjelovicZ12}).
Recent advances in deep-learning-based object detection have lifted this approach
to a higher level, by assigning object labels to bounding boxes 
(e.g., \cite{hoffman2014lsda,DBLP:conf/cvpr/RedmonDGF16,DBLP:journals/pami/RenHG017,DBLP:journals/pami/VinyalsTBE17}). However, these labels are limited to concrete object classes
(e.g., \textit{truck, SUV, Toyota Yaris Hybrid 2016}, etc.),
often trained (only) on (subsets of) the ca. 20,000 classes 
of ImageNet~\cite{DBLP:conf/cvpr/DengDSLL009}.
Thus, they miss out on generalizations (e.g., \textit{vehicle}) and abstractions
(e.g., \textit{transportation, traffic jam, rush hour}).
Fig~\ref{existingIssues} (a) shows 
the top-confidence visual labels by LSDA~\cite{hoffman2014lsda} 
for an example case of incorrect labels.

TBIR retrieves images by textual matches between user query and manually assigned image tags
(e.g., from collaborative communities such as Flickr).
While some of the semantic gap in CBIR is reduced in TBIR, the performance of TBIR often suffers from 
incomplete and ambiguous tags \cite{halpin2007complex}.
Figure~\ref{existingIssues} (b)
illustrates this point: there is only a single tag {\em happiness} and none for the
concrete objects in the image.
For the big search engines, one way of overcoming this bottleneck
is to exploit query-and-click logs (e.g., \cite{DBLP:conf/sigir/CraswellS07,
DBLP:conf/icmcs/HuaL14,
DBLP:journals/tip/WuLSYZRZ16}).
The query keyword(s) associated with a click can be treated as label(s)
for the clicked image. However, this method 
crucially relies
on the labels to appear in (sufficiently many) queries (or, traditionally, 
salient text surrounding the image).
\cite{li2016socializing} gives a
survey on TBIR and tag assignment and refinement.

Recently, the gap between the two image search paradigms
is narrowing. TBIR-style tags 
inferred from query-and-click logs
can be used to train a deep-learning network for more informative labels towards
better CBIR. 
Also, crowdsourcing could be a way towards more semantic labels (e.g., \cite{DBLP:journals/ftcgv/KovashkaRFG16}), for example, to capture human activities or emotions 
(e.g., \cite{DBLP:conf/cvpr/DivvalaFG14,DBLP:conf/iccv/GkioxariGM15,DBLP:conf/cvpr/RamanathanLDHLG15,kostiemotion:CVPR2017}).
Nevertheless, there are still major shortcomings in 
the state-of-the-art.

This paper addresses the outlined limitations.
The goal is to automatically annotate images
with semantically informative tags, including
generalizations and abstractions and also cleaning out
noisy labels for concrete objects.\\

\vspace{-0.5em}
\noindent
\textbf{Approach and Contribution:}
We leverage state-of-the-art CBIR by 
considering the visual tags of an existing
object detection tool (LSDA~\cite{hoffman2014lsda}
in our experiments)
as a starting point. Note that there are
multiple labels for each bounding box with
varying confidence scores, and our goal is
to compute the most informative labels for
the entire image.
We impose a constrained optimization on
these initial labels, in order to enforce
their semantic coherence. 
We also consider labels that are visually
similar to the detected ones, to compensate
for omissions.
In addition, we
utilize lexical and commonsense knowledge
to generate candidate labels for 
generalizations (hypernyms from 
WordNet \cite{Miller:1995:WLD:219717.219748}) 
and abstractions
(concepts from ConceptNet 
\cite{DBLP:conf/lrec/SpeerH12}).
So we both refine and expand the initial labels.
The joint inference on the entire label space
is modeled as an optimization problem,
solved by an integer linear program
(viz. using the Gurobi ILP solver).
Figure~\ref{goodExamples} shows examples
for the input labels from the deep-learning-based
visual object detection (left column) and
the output labels that VISIR computes (right column).
The labels from LSDA illustrate a clear 
semantic incoherence for these specific examples. VISIR labels are
coherent, adds generalizations (in blue)
and abstractions (in green). Incorrect
labels are marked red.
Although our work aligns with the existing 
TBIR research
on social tagging and tag refinement, there are
key
differences.\\

\vspace{-0.9em}
\noindent
{\em Granularity:}  Our starting point is
labels for bounding boxes, whereas user-provided
tags refer to an entire image.\\
{\em Cardinality:}  The number of bounding boxes
in one image can be quite large. Moreover,
object detectors usually produce a long list
of varying-confidence labels for each
bounding box.\\
{\em Noise:} As a result, many of the 
visual candidate labels in our approach
are of mixed quality, whereas traditional
social tagging typically has few but
trusted annotations per image.\\
For these factors, our notion of tag
refinement is unlike the one in prior work.
Therefore, we refer to our task as
{\em Visual Tag Refinement}.\\

\vspace{-0.5em}
\noindent
{\em Visual Tag Refinement} can be broken down 
into three sub-tasks, for which this paper
provides effective solutions:\\
1) elimination of incoherent tags~\footnote{``tag'' and ``label'' are used interchangeably in the paper with the same meaning.}
among the initial visual labels,\\
2) expansion of the tag space by adding visually similar tags missed by object detectors, and adding candidate tags for generalization and abstraction,
3) joint inference on the enriched
tag space, by integer linear program.
\section{Related Work}
\label{relatedWork}

\noindent
{\bf Automatic Image Annotation:} 
Early work on this problem generated tags only for an entire image
(or a single image region), but did so one class at a time
(e.g., \cite{akbas2007automatic}).
More recent methods support labeling multiple objects in the same image.
One such approach, WSABIE~\cite{weston2011wsabie}, performs k-nearest-neighbor classification on embeddings of words and features to scale to many classes. 

State-of-the-art work on object detection 
addresses both recognizing object bounding boxes and tagging them with their class labels.
Such work makes heavy use of deep learning (especially convolutional neural networks or CNNs).
Prominent representatives are LSDA~\cite{hoffman2014lsda}, and Faster R-CNN.
The latter \cite{DBLP:journals/pami/RenHG017}
improves the speed of object detection
by incorporating a Region Proposal Network (RPN).
Major emphasis in this line of Computer Vision work has been on
coping with small, partly occluded and poorly illuminated objects.
In contrast, the emphasis of VISIR is on the semantic coherence
between objects and jointly modeling the uncertainty of
candidate labels. Instead of speed, we optimize for higher labeling quality. 
VISIR is agnostic of the underlying object detector;
it is straightforward to plug in a different tool.

Context plays an important role in computer vision \cite{divvala2009}, and context-based object detectors were popular before the 
success of CNNs. 
These methods consider 
local context \cite{carreira2011}, global context \cite{rabinovich2007} or their combination \cite{yao2012}. 
With the advent of CNNs, 
the focus shifted to representation learning and improving detection speed. However, contextual information is gaining renewed attention. 
The state-of-the-art method
\cite{DBLP:conf/cvpr/WangYMHHX16} is based on a CNN-RNN combination, where the recurrent neural network (RNN) captures label dependencies. 
YOLO~\cite{DBLP:conf/cvpr/RedmonDGF16,DBLP:journals/corr/RedmonF16}
unifies learning with global context into a single neural network for the entire image. It exploits a word tree derived from the WordNet whose leaf nodes appear in ImageNet.
In our experiments, we use YOLO as a baseline for context-aware
object detection.
All learning-based methods crucially rely on extensive
training data.\\

\vspace{-0.5em}
\noindent
{\bf Social Tagging:}
TBIR has its origin in community-based social tagging of
images (e.g., Flickr), web pages or publications (e.g., Bibsonomy).
Crowdsourcing to compile large training collections
can be seen as a variant of this kind of user-provided tagging.
There is ample research in this area~\cite{DBLP:journals/tmm/LiSW09,DBLP:reference/rsh/MarinhoNSJHSS11,DBLP:journals/tip/GaoWZSLW13}, 
especially on learning tag recommendations.
Our task of visual tag refinement differs from social tagging
substantially. The CBIR-based tags that we start with,
label individual objects instead of assigning tags to the overall image. Multiple candidate tags per bounding box also lead to a dense tag space in contrast to sparse user tags. 
Finally, the large number of varying-granularity and
varying-confidence tags per image entails a much higher
degree of noise in the label space, whereas social tags
are usually considered trusted.\\

\vspace{-0.5em}
\noindent
{\bf Tag Refinement:} The problem of tag refinement aims at removing noisy tags from images while adding more relevant ones~\cite{li2016socializing}. This line of work appears in the literature also as tag completion~\cite{wu2013tag, feng2014image} or image re-tagging~\cite{liu2010retagging, liu2011image}. 
Most of this work uses only nouns as tags and disregards word ambiguity. Background knowledge, such as WordNet 
synonymy sets and other lexical relations,
is rarely used. Word categories that are vital for
denoting abstractions, namely, adjectives, verbs and verbal phrases,
are out of scope.
Moreover a common assumption is that
visually similar images are semantically similar, meaning that they 
should have similar tags. This assumption is often invalid. 
This body of work employs 
a variety of methods, including metric learning~\cite{guillaumin2009tagprop},
matrix completion~\cite{wu2013tag, feng2014image,zhu2010image}, 
latent topic models~\cite{xu2009tag}, 
and more.\\

\vspace{-0.5em}
\noindent
{\bf Commonsense Knowledge for Image Retrieval:} The first work
on image retrieval with commonsense knowledge (CSK)~\cite{liu2002robust} exploited
the Open Mind Commonsense Knowledgebase~\cite{singh2002open}, 
a small knowledge base with
simple properties of concepts for concept expansion and for
activation spreading.
Since then, much more comprehensive CSK knowledge bases
have been constructed, most notably, 
ConceptNet \cite{DBLP:conf/lrec/SpeerH12} and WebChild \cite{tandon2014webchild}. 
However, such background knowledge has not been used by
modern object detectors.
A notable exception addresses emotions invoked by images,
and tags objects with sentiment-bearing adjectives \cite{chen2014object}.
However, this work is limited to a small label space. A recent framework for image search~\cite{DBLP:conf/akbc/ChowdhuryTW16} uses CSK extracted through OpenIE for query expansion.
\vspace{-0.5cm}
\section{Model and Methodology}
\label{visualTagRefinement}

We define the problem of {\em Visual Tag Refinement} as the tasks of --\\
1) cleaning noisy object tags from low-level image features\\
2) enriching existing detections by adding additional relevant tags\\
3) abstracting from 
concrete objects towards a more conceptual space. We first present a framework for the proposed problem, followed by the description of its individual components. We then present the optimization model to solve the problem.

\vspace{-1.1em}
\subsection{Framework for Visual Tag Refinement}
\label{VISIR_framework}

We consider an image $x$ with multiple bounding boxes $x_1, x_2,...x_k$. 
Each bounding box $x_i$ has labels for detected physical objects 
along with detection confidence scores. 
The values of these labels and scores are outputs of 
an off-the-shelf object detection tool, e.g., LSDA~\cite{hoffman2014lsda}.  We define three different label spaces, candidates from which would be associated either with bounding boxes of an image or would globally add semantics to it:
\squishlist
\item A space of all possible object labels detectable by an underlying object detection tool is denoted by $CL$
(for Concrete-object Labels).
For example, the LSDA tool~\cite{hoffman2014lsda} uses ImageNet~\cite{DBLP:conf/cvpr/DengDSLL009} object classes, which are also leaf-level synsets of WordNet~\cite{Miller:1995:WLD:219717.219748}. 
There can be two sets of such classes for a bounding box $x_i$ of image $x$: 
$cl_i$ constitute those labels originally detected from low-level image features, 
$cl_i'$ constitute undetected labels visually similar to those in $cl_i$.

\item A space of extended labels is denoted by $XL$. For each image $x$ and bounding box $x_i$, a subset from this space, 
$xl_i$, contains additional label candidates that generalize the classes in $cl_i$ and $cl_i'$. For example, ``ant'' $\in cl_i \rightarrow$ ``insect'' $\in xl_i$. Adding generalized terms to the label space serves a dual purpose - overcoming the training bias of the object detection tools, and broadening the label space for greater web visibility. We discuss more on the issue of training bias in section~\ref{discussions}.

$CL$ and $XL$ contain labels signifying visual objects in an image. Hence we call the super set of such labels ``visual labels'' $VL$; $VL = CL \cup XL$.

\item A space of abstract labels (utilities, emotions, themes) is denoted by $AL$. This constitutes abstract concepts associated with visual objects derived from commonsense knowledge bases. 
For example, \textit{fragrant} $\in AL$ from the ConceptNet~\cite{liu2004conceptnet} clause \texttt{hasProperty(flower, fragrant)}.
\squishend

\noindent
An image $x$ can hence be described by three sets of labels - 
a set of deep-learning based class labels $cl \cup cl' \in CL$, 
a set of extended labels $xl \in XL$, and a set of abstract labels $al \in AL$.
Further, we define three different scores that act as edges between nodes of the above spaces:
\squishlist
\item Visual Similarity $vsim(l_j, l_k)$ for $l_j, l_k \in CL$
\item Semantic Relatedness $srel(l_j,l_k)$  for $l_j, l_k \in VL$
\item Abstraction Confidence $aconf(l_j, al_n)$ for $al_n \in AL$ and $l_j \in VL$
\squishend

We present the visual tag refinement problem in terms of three sub-problems:
\squishlist
\item The noisy tag problem - for each image $x$ and bounding box $x_i$ infer which of the labels in $cl_i \in CL$ should be accepted. We eliminate those labels which are not coherent with the other bounding box detections in the image. For example, in Figure~\ref{goodExamples} image 5 we eliminate the detection \textit{cucumber} since it is not semantically related to the other labels \textit{snake} and \textit{green mamba}.
\item The incomplete tag problem - for each image $x$ and bounding box $x_i$ infer which of the labels in 
$cl_i' \in CL$ and $xl_i \in XL$ should be additionally associated with the bounding box.
\item The abstraction tag problem - infer which of $al \in AL$ should be globally associated with image $x$.
\squishend
We solve these problems jointly and retain the most confident hypothesis for each bounding box relative to the others as well as a global hypothesis toward tag abstraction in an image. Hence, we predict a set of plausible labels $L_x \in CL \cup XL \cup AL$ for an image $x$.

\vspace{-0.5em}
\subsection{Visual Similarity (or ``Confusability'')}

Deep-learning based tools using low-level image features to predict the object classes can confuse one object to be another. We consider two labels to be visually similar if they occur as candidates in $cl_i$ for the same bounding box $x_i$. We collect evidence of such visual similarity from low-level image features, in particular, from object detection results of LSDA~\cite{hoffman2014lsda}. 
We define the visual similarity between two labels $l_j$ and $l_k$ by a Jaccard-style similarity measure as shown in Equation~\ref{vsim_jaccard}.
In this similarity measure, if labels $l_j$ and $l_k$ always appear together as candidates for the same bounding box, and never with any other labels, then they are considered highly visually similar, $vsim(l_j, l_k)=1$. If labels $l_j$ and $l_k$ never appear together, one label is never confused by the tool to be another; in this case $vsim(l_j, l_k)=0$, meaning $l_j$ and $l_k$ are not visually similar. Given that the initial object detections from low level image features are noisy in itself, this evidence would also contain noise. However, it is expected that the evidence will hold when it is computed over a large dataset. 
\begin{multline}
    \label{vsim_jaccard}
    vsim(l_j, l_k)=\\ 
    \dfrac{\sum\limits_{i: l_j, l_k \in cl_i}(conf_{BB}(x_i,l_j) + conf_{BB}(x_i,l_k))}
    {\sum\limits_{i: l_j \in cl_i}(conf_{BB}(x_i,l_j)) + \sum\limits_{i: l_k \in cl_i}(conf_{BB}(x_i,l_k))}
\end{multline}
We can refer to this measure also as ``confusability'' since the object detection tool confuses one object to be another based on similar low-level visual features.

\vspace{-0.5em}
\subsection{Semantic Relatedness}
\label{semRel}
Semantic Relatedness between two concepts signifies their conceptual similarity. 
Our model uses this measure to establish the contextual coherence between labels of different bounding boxes.
The relatedness between two labels $l_j$ and $l_k$ is defined as a weighted linear combination of their cosine similarity from word embeddings and their spatial co-location confidence.
\begin{multline}
    \label{srel}
    srel(l_j, l_k) = \delta cosine(l_j, l_k) + (1-\delta) coloc(l_j, l_k)\\
                                                for\ l_j, l_k \in VL
\end{multline}

\noindent
{\bf Word Embeddings: }
To improve the contextual coherence between object labels in images, 
the context of words needs to be captured. 
We utilize vector space word embeddings for this purpose.  
A word2vec \cite{mikolov2014word2vec} model
is trained from manually annotated image descriptions from 
a large set of image captions, as described later in more details. 
The cosine similarity between two labels -- $cosine(., .)$  in Equation~\ref{srel} -- is calculated from their respective word vectors.\\

\vspace{-0.8em}
\noindent
{\bf Spatial Co-location:}
Spatial relationships between concepts carry an important evidence of relatedness. 
For example, an ``apple'' and a ``table'' are related concepts since they occur in 
close spatial proximity. Similarly, a ``tennis racket'' and a ``lemon'' are unrelated. 
$coloc(., .)$ in Equation~\ref{srel} is a frequency-based co-location score mined from manual annotations of image labels.

\vspace{-0.5em}
\subsection{Concept Generalization}
A hypernym is a superordinate of a concept. 
In other words, a concept is an instantiation of its hypernym. 
For example, \textit{fruit} is a hypernym for \textit{apple}, 
i.e., \textit{apple} IsA \textit{fruit}. 
WordNet~\cite{Miller:1995:WLD:219717.219748} provides a hierarchy of concepts 
and their hypernyms which we leverage to generalize our object classes. 
WordNet also reports different meanings (senses) of a concept; 
for example a \textit{punching bag} is (\textit{a person on whom another person vents their anger}) 
or (\textit{an inflated ball or bag that is suspended and punched for training in boxing}), 
leading to very different hypernymy trees. 
For this reason, we map our object classes from ImageNet into their correct WordNet sense number, 
followed by traversing their hypernymy tree up to a certain level. 
This yields a cleaner generalization. 
Further more, to avoid exotic words among the hypernyms, 
we use their approximate Google result counts and prune out those below a threshold. 
Hence for the concept \textit{ant} we retain the hypernym \textit{insect} 
and prune the hypernym \textit{hymenopteran}. 
Following this heuristics, we assign 1 to 3 hypernyms per object class.

\vspace{-0.5em}
\subsection{Concept Abstraction}
To introduce human factors like commonsense and subjective perception, 
we incorporate abstract words and phrases associated with visual concepts of an image. 
For example an \textit{accordion} is ``used to'' \textit{make music}. 
We consider two relations from ConceptNet 5~\cite{liu2004conceptnet} 
for assigning the abstract labels - \textit{usedFor}, and \textit{hasProperty}. 
Some example of assigned abstract labels/phrases (in green) can be found in Figure~\ref{goodExamples}. 
Abstract concepts which are assigned to images have high abstraction confidence. Abstraction confidence of a concept/phrase is defined as the joint semantic relatedness of the phrase and the refined visual labels of the image.

\vspace{-0.5em}
\subsection{Tag Refinement Modeled as an ILP}
We cast the multi-label visual tag refinement problem into an Integer Linear Program (ILP) optimization with the following definitions.
We choose ILP as it is a very expressive framework
for modeling constrained optimization 
(more powerful than probabilistic graphical models),
and at the same time comes with very mature and efficient
solvers like Gurobi (\url{http://gurobi.com}).
Some tools for probabilistic graphical models
even use ILP for efficient MAP inference.

\vspace{0.4em}
\noindent
Given an image $x$, with bounding boxes $x_1, x_2,...$, 
it has three sets of visual labels: 
$cl_i$ (initial bounding box labels), 
$cl_i'$ (labels visually similar to the original detections), 
and $xl_i$ (hypernyms of labels in $cl_i \cup cl_i'$). 
The set $vl_i = cl_i \cup cl_i' \cup xl_i$ 
constitutes all visual labels which are candidates for bounding box $x_i$. 
The image would also be assigned abstract labels $al_1, al_2, ...$ globally. 
We thus introduce 0-1 decision variables:

\noindent
$X_{ij} = 1$ if $x_i$ should indeed have visual label 
$vl_{j}$, 0 otherwise\\
$Y_j = 1$ if $x$ should indeed have abstract label $al_j$, 0 otherwise\\
$Z_{ijmk} = 1$ if $X_{ij}=1$ and $X_{mk}=1$, 0 otherwise \\
$W_{ijk} = 1$ if $X_{ij}=1$ and $Y_{k} = 1$, 0 otherwise

\noindent
Decision variables $Z_{ijmk}$ and $W_{ijk}$ emphasise pair-wise coherence 
between two visual labels and between a visual and an abstract label respectively.

\vspace{0.4em}
\noindent
{\bf Objective:}
Select labels for $x$ and its bounding boxes which maximizes
a weighted sum of evidence and coherence --
\begin{multline}
\label{eq:objective}
    max \bigg[ \alpha \sum\limits_{i,j}\big(vconf(x_i,l_{j}) + \kappa gconf(x_i, l_{j})\big) X_{ij} + \\
    \beta \sum\limits_{i,m}\sum\limits_{\substack{l_j \in vl_i \\ l_k \in vl_m} } srel(l_{j},l_{k}) Z_{ijmk} +\\
    \gamma \sum\limits_{l_j \in VL}\sum\limits_k aconf(l_j, al_k)\sum\limits_i W_{ijk} \bigg]
\end{multline}
with hyper-parameters $\alpha$, $\beta$, $\gamma$, $\kappa$.

\vspace{0.6em}
\noindent
For each $l \in CL$, we define set $S(l) \subseteq CL$
of labels visually similar to $l$. 
$vsim(l,l')=0$ if $l' \notin S(l)$. 
Recall the definition of $vsim(.,,)$ from Equation~\ref{vsim_jaccard}.

\noindent
\textit{Visual Confidence}, the confidence with which a visual label should be associated with an image is defined as:
\begin{align}
vconf(x_i, l_j) &= conf_{BB}(x_i, l_j) \text{ if } l_j \in cl_i\\
                &= \sum\limits_{l\in cl_i}conf_{BB}(x_i,l)vsim(l,l_j) \text{ if } l_j \in cl_i'/cl_i
\end{align}
\noindent
Here, a high confident original detection adds significant weight to the objective function, 
hence increasing the chances of its retention. Similarly, the weight of a label visually similar 
to multiple original labels is boosted. Also, labels visually similar to only one low confident 
original label is assigned less importance.

\noindent
For $l \in CL$ we define a set $H(l) \in  XL$ of hypernyms of $l$.
The \textit{Generalization Confidence} of a label $l_j$ in bounding box 
$x_i$ is defined in terms of the semantic relatedness between the label and its hypernym.
\begin{flalign}
gconf(x_i,l_j) &= \sum\limits_{l:l_j\in H(l)} srel(l_j, l)\ if\ l_j \in xl_i\\
               &= 0\ if\ l_j \in \{cl_i \cup cl_i'\}
\end{flalign}

\noindent
\textit{Abstraction Confidence} $aconf(.,.)$ of a label $l_j$ and an abstract concept $al_k$ 
is defined as their semantic relatedness, weighted by the score of the
assertion containing the abstract concept in ConceptNet~\cite{liu2004conceptnet}.
For example, \texttt{hasProperty(baby, newborn)} has a score of 10.17 in ConceptNet.
We name this score $CNet(al_k)$. 
\begin{equation}
aconf(l_j, al_k) = CNet(al_k) * srel(l_j, al_k)
\end{equation}

\vspace{0.2cm}
\noindent
{\bf Constraints:}\\
$\sum_j X_{ij} <= 1$ : for each bounding box $x_i$ there can be at most one visual label ($\in VL$)\\
$\sum_j Y_j <= 5$ : one image $x$ can have at most five abstract labels ($\in AL$)\\
\begin{equation*}
\begin{rcases}
(1-Z_{ijmk}) <= (1-X_{ij}) + (1-X_{mk})\\
Z_{ijmk} <= X_{ij}\\
Z_{ijmk} <= X_{mk}
\end{rcases}
\text{\parbox{2.7cm}{Pair-wise mutual \\coherence between visual labels}}
\end{equation*}
\begin{equation*}
\begin{rcases}
(1-W_{ijk}) <= (1-X_{ij}) + (1-Y_k)\\
W_{ijk} <= X_{ij}\\
W_{ijk} <= Y_k
\end{rcases}
\text{\parbox{2.7cm}{Pair-wise mutual \\coherence between visual and abstract \\labels}}
\end{equation*}

\vspace{0.4cm}
The final set of visual and abstract labels per image are expected to be highly coherent. 
This is validated in Section~\ref{experiments}.

\section{Data Sets and Tools}
\label{dataTools}

In this section, we present the image data sets as well as the criteria and heuristics we follow to mine the various background knowledge utilized in our optimization model.\\ 

\vspace{-0.5em}
\noindent
\textbf{ImageNet Object Classes:}
LSDA~\cite{hoffman2014lsda} is used to get the initial visual object labels from low-level image features.  
The LSDA tool has been trained on 7604 leaf-level nodes of ImageNet
\cite{DBLP:conf/cvpr/DengDSLL009}. 
Most of these object classes are exotic concepts which rarely occur in everyday images. 
Examples include scientific names of flora and fauna -- \textit{interior live oak, Quercus wislizenii, 
American white oak, Quercus alba}, 
and obscure terms -- \textit{pannikin, reliquary, lacrosse}. 
We prune those exotic classes by thresholding on
their Google and Flickr search result counts.
Some object class names are ambiguous where 
two senses of the same word from WordNet have been included.
We consider only the most common sense.
We work with the most frequent 1000 object classes obtained after pruning~\footnote{The full list is available at \url{http://people.mpi-inf.mpg.de/~sreyasi/visTagRef/1000classes_names.txt}}.\\

\vspace{-0.5em}
\noindent
\textbf{WordNet Hypernyms: }
For the ImageNet object classes described above, we traverse the WordNet hypernymy tree of the associated sense up to level three. 
We restrict the traversal level to avoid too much generalization -- 
for example, \textit{person} generalizing to \textit{organism}.
We prune out hypernyms with Google and Flickr result counts below a threshold. 
By considering the hypernyms of the 1000 ImageNet 
object classes mentioned above, we add ~800 new visual labels to the model.\\
The ImageNet object classes and the WordNet hypernyms together constitute the \textbf{Visual Labels} of VISIR.\\

\vspace{-0.5em}
\noindent
\textbf{Abstract Labels: }
Commonsense knowledge (CSK) assertions from ConceptNet \cite{DBLP:conf/lrec/SpeerH12} contribute to concept abstraction in VISIR. 
For example, in Figure~\ref{goodExamples}, the abstract concept \textit{poisonous} is added to the labels
of the fifth image. 
ConceptNet is a crowd-sourced knowledge base where 
most assertions have the default confidence score of 1.0 
(as they were stated only by one person). 
Only popular statements like \texttt{hasProperty(apple, red fruit)} 
are stated by multiple people, hence raising the confidence score significantly. 
Certain assertions have contradictory scores -- for example, \texttt{usedFor(acne medicine, clear skin)} 
appears twice, with scores 1.0 and  -1.0. 
This happens when someone down-votes a statement. 
Using such indistinctive scores in VISIR 
would be uninformative. 
We therefore use the joint semantic relatedness of the assertion and visual labels of an image, 
weighted by the ConceptNet score (only positive scores), as the abstraction confidence.\\

\vspace{-0.5em}
\noindent
\textbf{Visual Similarity:}
The visual similarity or ``confusability'' scores (Equation~\ref{vsim_jaccard}) 
are mined from object detection results (from low-level image features) 
over 1 million images from the following data sets that are popularly employed in the computer vision community:
Flickr 30k~\cite{young2014image}, 
Pascal Sentence Dataset~\cite{rashtchian2010collecting}, 
SBU Image Dataset~\cite{Ordonez:2011:im2text}, 
MS-COCO~\cite{lin2014microsoft}.
All these data sets have collections of Flickr images not pertaining to any particular domain. 
For each detected bounding box, LSDA provides a confidence score distribution over 7604 object classes (leaf nodes in ImageNet). 
Only predictions with a positive confidence score are considered as candidates for a bounding box. 
An object class pair appearing as candidates for the same bounding box are considered as visually similar. 
Table~\ref{visSim} shows few examples of visually similar object class pairs -- 
\textit{mail train} and \textit{commuter train} are confused 91\% times whereas \textit{diaper} and \textit{plaster cast} are confused 18\% times.\\
\begin{table}[ht]
\centering
\caption{Object class pairs and visual similarity scores \vspace{-0.5em}}
\label{visSim}
\begin{tabularx}{\columnwidth}{XXX}
\hline
object1       & object2        & visual similarity \\
\hline
mail train    & commuter train & 0.91              \\
cattle        & horse          & 0.76              \\
soccer ball   & kite baloon    & 0.26              \\
Red Delicious & bowling ball   & 0.21              \\
diaper        & plaster cast   & 0.18   		    \\    
bicycle pump  & mascara        & 0.17              
\end{tabularx}
\end{table}

\vspace{-0.5em}
\noindent
\textbf{Spatial Co-location: }
Spatial co-location scores between different object classes are mined from ground truth annotations 
of the detection challenge (DET) of ImageNet ILSVRC 2015~\cite{ILSVRC15}. 
We consider two objects to be spatially co-located only if they are tagged in the same image.  
For simplicity, we do not consider the physical distance between the bounding boxes of the tagged object classes. 
A frequency-based co-location score is assigned to pairs of object classes based on evidence over the train set of ILSVRC DET. 
We find spatial co-location data for 200 object classes (since the detection challenge only considers 200 object classes). 
The top few frequently co-located objects are: 
\textit{(person, microphone), (table, chair), (person, sunglasses), (person, table), (person, chair)}. 
A general observation would be that the image collection in ILSVRC DET has a high occurrence of \textit{person}.
\section{Experiments and Results}
\label{experiments}
We analyze and compare the results that VISIR produces with that of two baselines: LSDA~\footnote{\url{http://lsda.berkeleyvision.org/}} and YOLO~\footnote{\url{https://pjreddie.com/darknet/yolo/}}.
The performances of LSDA, YOLO, and VISIR are compared on the basis of precision, recall and F1-score measures.

\subsection{Setup}
As discussed in Section~\ref{VISIR_framework}, we operate with
three kinds of labels:
visual class labels from ImageNet ($CL$),
their generalizations ($XL$) which consist of WordNet hypernyms of labels $\in CL$,
and abstract labels ($AL$) from commonsense knowledge.
We evaluate the methods with respect to three different label spaces (as the combination of three
types of labels):
$CL$, $CL+XL$, and $CL+XL+AL$.
LSDA and YOLO operate only on CL, while VISIR has three variants
(configuring it for the above combinations of label spaces).
Each system is given a label budget of 5 tags
per image. For VISIR, this is enforced by
an ILP constraint; for the two baselines, 
we use their confidence scores to pick the top-5.\\

\vspace{-0.5em}
\noindent
\textbf{Hyper-parameter Tuning: }
To tune the hyper-parameters for Equation~\ref{eq:objective} 
we use the annotations of the training image set of ILSVRC DET.
We also extend this set by adding the hypernyms of the 
ground-truth labels as correct labels.
A randomized search is used to tune the hyperparameters.\\

\vspace{-0.5em}
\noindent
\textbf{User Evaluation:}
Besides establishing semantic coherence among concrete object labels, VISIR applies concept generalization and abstraction. 
For modern benchmark datasets like ILSVRC 2015 DET, such enriched labeling does not exist so far.
Therefore, in order to evaluate VISIR and compare to baselines,
we construct a labeled image dataset by collecting human judgments 
about correctness of labels as discussed below.

For each label space, $CL$, $CL+XL$, and $CL+XL+AL$, the union of the labels produced by each method forms the set of result labels for an image. This result pool is evaluated by human annotators. Judges determined whether each label is appropriate and informative for an image. Instead of a binary assessments, annotators are asked to
grade each label in the pool with 0, 1, or 2 -- 0 corresponding to incorrect
labels, 2 corresponding to highly relevant labels. We
gather user judgments for the three label pools
(corresponding to the label spaces) separately. This produces three different sets of graded labels per image. Users are not informed about the nature of the label pools.
For each label pool we collect responses from at least 5 judges. The final assessment is determined by the majority of the judges (e.g., at least 3 out of 5 need to assert that a label is good).
\\

\vspace{-0.5em}
\noindent
\textbf{Selection of Test Images: }
A major goal of this work is to make the refined labels more  coherent or semantically related. Hence, we focus on the case where the deep-learning-based object detection tools produce contextually incoherent results. For the user evaluation, we collect a set of images with a reasonable context -- those that have 3-7 detected bounding boxes and with LSDA labels having a semantic relatedness score less than 0.1. Such 100 images are collected from the ILSVRC 2015 DET Val image set.

\vspace{-0.6em}
\subsection{Model Performance}
Precision is estimated as the fraction of ``good labels''
detected, where a ``good label'' is one considered relevant by
the majority of the human judges.
We assess the recall per method as the number of labels picked from the good labels
in the pool of labels generated by all three methods. 
The recall is artificially restricted because 
the label pool may contain more good labels than the label budget of the
method. For example, if the label budget per method is set to 5,
even if all 5 labels of a method are good, the recall for a pool with
8 good labels would only be 5/8. 
However, it is a fair notion across the different methods.\\

\vspace{-0.5em}
\noindent
\textbf{Relaxed vs Conservative Assessments: }
According to the evaluation design, labels graded 1 are either
inconspicuous, or less relevant to the image than labels graded 2. In order to identify the ``good labels'' in a label pool, we
define two methods of assessment: \textit{Relaxed Assessment} considers all labels graded 1 or 2 as
correct. \textit{Conservative Assessment} considers only those labels graded 2 as correct,
resulting in a stricter setup.
The three graded label pools from the user evaluations
naturally have labels in common.\\

\vspace{-0.5em}
\noindent
\textbf{Performance Results: }
Tables~\ref{results_CL} through~\ref{results_CL+XL+AL} compare the three methods
for the three different label pools -- $CL$, $CL+XL$, $CL+XL+AL$ -- with conservative assessment. 
For $CL$, there is no real improvement over LSDA, but we see that for $CL+XL$ and $CL+AL+XL$ VISIR adds
a good number of semantically informative labels
and improves on the two baselines in terms of
both precision and recall.

We also test VISIR's performance with a tighter constraint on choosing the number of bounding boxes per image, by setting the label budget to 80\% of all bounding boxes received as input.  This variant, which we refer to as VISIR\textsuperscript{*}, aims to filter out more noise in the output of the deep-learning-based object detections. 
Naturally, 
VISIR\textsuperscript{*}-CL would have higher precision than VISIR-CL while sacrificing on recall.
VISIR\textsuperscript{*}-CL improves further on precision and F1-score because it is able to
eliminate some of the initial noise the LSDA detections bring in.
For pools $CL+XL$ and $CL+XL+AL$,
VISIR\textsuperscript{*} has higher precision than VISIR, but slighly loses in recall.

\begin{table}[ht]
\centering
\caption{Pool CL: Conservative Assessment \vspace{-0.7em}}
\label{results_CL}
\begin{tabularx}{\columnwidth}{bsss}
\hline
System & Precision & Recall & F1-score\\
\hline
LSDA            &	0.51    &	0.86    &	0.64\\
YOLO	        &   0.49	&   0.56	&   0.52\\
\hline
VISIR-CL        &	0.51	&   0.86	&   0.64\\
VISIR\textsuperscript{*}-CL&   \textbf{0.57}    &   0.81    &   \textbf{0.67}\\
\end{tabularx}
\end{table}

\begin{table}[ht]
\centering
\caption{Pool CL+XL: Conservative Assessment \vspace{-0.7em}}
\label{results_CL+XL}
\begin{tabularx}{\columnwidth}{bsss}
\hline
System & Precision & Recall & F1-score\\
\hline
LSDA            &	0.52    &	0.81    &	0.63\\
YOLO	        &   0.49	&   0.51	&   0.50\\
\hline
VISIR-CL+XL        &	\textbf{0.54}	&   \textbf{0.82}	&   \textbf{0.65}\\
VISIR\textsuperscript{*}-CL+XL&   \textbf{0.60}    &   0.76    &   \textbf{0.67}\\
\end{tabularx}
\end{table}

\begin{table}[ht]
\centering
\caption{Pool CL+XL+AL: Conservative Assessment \vspace{-0.7em}}
\label{results_CL+XL+AL}
\begin{tabularx}{\columnwidth}{bsss}
\hline
System & Precision & Recall & F1-score\\
\hline
LSDA            &	0.49    &	0.35    &	0.41\\
YOLO	        &   0.52	&   0.23	&   0.32\\
\hline
VISIR-CL+XL+AL        &	\textbf{0.54}	&   \textbf{0.91}	&   \textbf{0.68}\\
VISIR\textsuperscript{*}-CL+XL+AL&   \textbf{0.56}    &   \textbf{0.89}    &   \textbf{0.69}\\
\end{tabularx}
\end{table}

\begin{figure*}
    \centering
    \begin{tabular}{|p{2.6cm}|p{2.4cm}|p{2.4cm}|p{2.4cm}|p{2.4cm}|p{2.4cm}|}
    \hline
            &  \vspace{0.01em}{\hspace{2.8em}LSDA} & \vspace{0.01em}{\hspace{2.8em}YOLO}& \vspace{0.01em}{\hspace{2.2em}VISIR-CL}& \vspace{0.01em}{\hspace{1.5em}VISIR-CL+XL}& \vspace{0.01em}{VISIR-CL+XL+AL}\\
    \hline
         \raisebox{-\totalheight}{\includegraphics[width=8.3em]{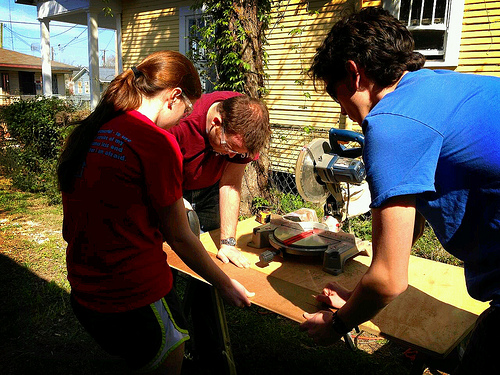}}\vspace{0.5em} 
         & 
         \vspace{0.01em}\makecell{person\\table\\{\color{ACMRed}motorcycle}}
         &
         \vspace{0.01em}\makecell{bench\\person\\{\color{ACMRed}bowl}}
         &
         \vspace{0.01em}\makecell{person\\table}
         &
         \vspace{0.01em}\makecell{person\\table}
         &
         \vspace{-1em}\makecell{\\person\\table\\man or woman\\furniture}\\ 
    \hline
         \raisebox{-\totalheight}{\includegraphics[width=8.3em]{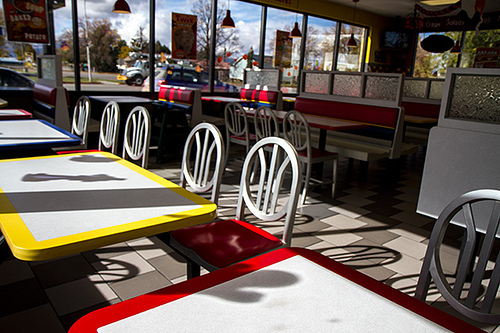}}\vspace{0.5em} 
         & 
         \vspace{0.01em}\makecell{table\\car} 
         & 
         \vspace{0.01em}\makecell{{\color{ACMRed}tow truck}\\bench\\car\\chair}
         &
         \vspace{0.01em}\makecell{table\\chair}
         &
         \vspace{0.01em}\makecell{table\\chair}
         &
         \vspace{-0.5em}\makecell{seat\\furniture\\chair\\{\color{ACMRed}flat}\\dining furniture\\table}\\ 
    \hline
         \raisebox{-\totalheight}{\includegraphics[width=8.3em]{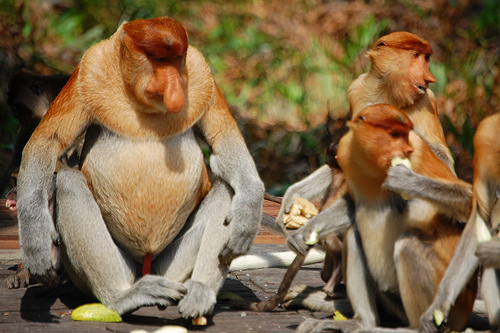}}\vspace{0.5em} 
         & 
         \vspace{0.01em}\makecell{monkey\\{\color{ACMRed}tennis ball}} 
         & 
         \vspace{0.01em}\makecell{{\color{ACMRed}bird}\\{\color{ACMRed}dog}}
         &
         \vspace{0.01em}\makecell{monkey}
         &
         \vspace{0.01em}\makecell{monkey}
         &
         \vspace{0.01em}\makecell{primate\\monkey\\{\color{ACMRed}orangutan}\\ape\\simian\\furry}\vspace{0.5em}\\
    \hline
         \raisebox{-\totalheight}{\includegraphics[width=8.3em]{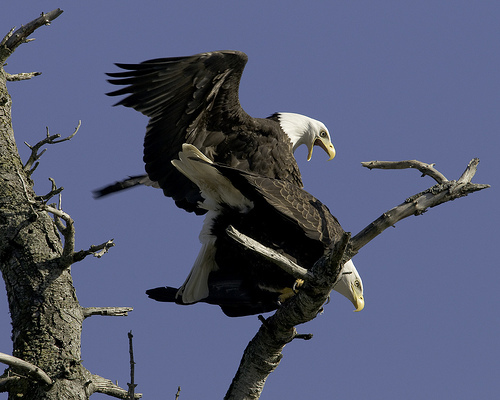}}\vspace{0.5em} 
         & 
         \vspace{0.01em}\makecell{bird\\{\color{ACMRed}hat with a}\\{\color{ACMRed}wide brim}} 
         & 
         \vspace{0.01em}\makecell{{\color{ACMRed}airplane}\\bird}
         &
         \vspace{0.01em}\makecell{bird}
         &
         \vspace{0.01em}\makecell{bird}
         &
         \vspace{0.01em}\makecell{bird\\avian\\flying animal}\\
    \hline
        \raisebox{-\totalheight}{\includegraphics[width=8.3em]{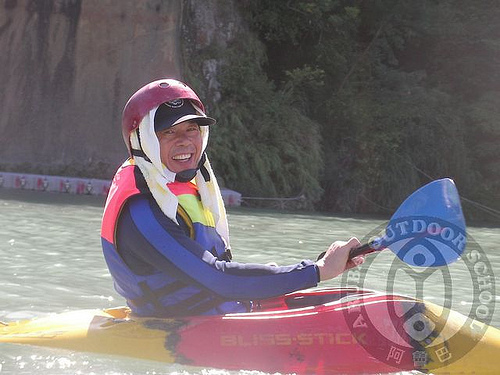}}\vspace{0.5em} 
        &
        \vspace{0.1em}\makecell{helmet\\person\\watercraft\\{\color{ACMRed}smelling bottle}\\{\color{ACMRed}bathing cap}\\{\color{ACMRed}record sleeve}\\{\color{ACMRed}impeller}}
        &
        \vspace{0.01em}\makecell{person} 
        &
        \vspace{0.01em}\makecell{{\color{ACMRed}bathing cap}\\{\color{ACMRed}bib}\\watercraft\\helmet\\person}
        &
        \vspace{0.01em}\makecell{{\color{ACMRed}bathing cap}\\fabric\\watercraft\\helmet\\person}
        & 
        \vspace{0.1em}\makecell{{\color{ACMRed}bathing cap}\\cloth, fabric\\watercraft\\helmet\\protective hat\\individual\\person}\vspace{0.5em}\\
    \hline
        \raisebox{-\totalheight}{\includegraphics[width=8.3em]{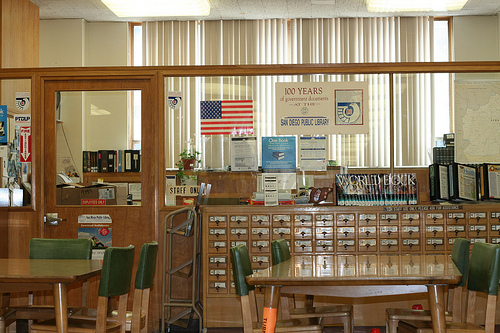}}\vspace{0.5em} 
         & 
         \vspace{0.01em}\makecell{table\\{\color{ACMRed}baby bed}\\{\color{ACMRed}swimming trunks}} 
         & 
         \vspace{0.01em}\makecell{chair} 
         &
         \vspace{0.01em}\makecell{chair\\table}
         &
         \vspace{0.01em}\makecell{chair\\table}
         &
         \vspace{0.01em}\makecell{seat\\furniture\\chair\\table\\{\color{ACMRed}flat}}\vspace{0.5em}\\
    \hline
    \end{tabular}
    \caption{\small Images with labels from LSDA, YOLO, and different configurations of VISIR}
    \label{anectodalEvidence-allMethods}
    \vspace{-0.5em}
\end{figure*}

Table~\ref{results_aggregate_relax} and
Table~\ref{results_aggregate_conserve} show the relaxed and
conservative assessments with respect to the 
combined pool (i.e., for all three label spaces
together)
of
good labels per image. 
It is natural that all methods perform better for the relaxed setting compared to that of the conservative assessment. However, the 
observation that VISIR's performance does not degrade much for the conservative assessment demonstrates
its high output quality and robustness.
Figure~\ref{anectodalEvidence-allMethods}
illustrates this by anecdotal examples with
the labels assigned by each of the competitors
(with good labels in black and bad ones in red).
In image 4, LSDA produces typically unrelated labels --
a \textit{monkey} and a 
\textit{tennis ball}. 
This contextual incoherence likely arises due to low level color features.
In contrast to LSDA, YOLO addresses the semantic coherence of the labels, 
however likely in expense of recall (for example in image 6). 
By necessitating semantic coherence among detected labels VISIR eliminates incoherent labels - for example,
VISIR removes \textit{motorcycle} from image 1, \textit{tennis ball}
from image 4, \textit{hat with a wide brim} from image 5 and so on. 

\begin{table}[ht]
\centering
\caption{Aggregate Pool: Relaxed Assessment \vspace{-0.7em}}
\label{results_aggregate_relax}
\begin{tabularx}{\columnwidth}{bsss}
\hline
System & Precision & Recall & F1-score\\
\hline
LSDA    &	0.55    &	0.30    &	0.39\\
YOLO	&   0.57	&   0.19	&   0.29\\
\hline
VISIR-CL&	\textbf{0.57}	&   0.28	&   0.38\\
VISIR-CL+XL&	\textbf{0.62}	&   0.30	&   \textbf{0.40}\\
VISIR-CL+XL+AL&	\textbf{0.71}&	\textbf{0.90}	&   \textbf{0.79}\\
\end{tabularx}
\end{table}

\begin{table}[ht]
\centering
\caption{Aggregate Pool: Conservative Assessment \vspace{-0.7em}}
\label{results_aggregate_conserve}
\begin{tabularx}{\columnwidth}{bsss}
\hline
System & Precision & Recall & F1-score\\
\hline
LSDA    &	0.49    &	0.35    &	0.41\\
YOLO	&   0.52	&   0.23	&   0.32\\
\hline
VISIR-CL&	\textbf{0.52}	&   0.34	&   0.41\\
VISIR-CL+XL&	\textbf{0.55}	&   \textbf{0.35}	&   \textbf{0.43}\\
VISIR-CL+XL+AL&	\textbf{0.54}&	\textbf{0.91}	&   \textbf{0.68}\\
\end{tabularx}
\end{table}

Table~\ref{solveIncompleteness} lists the new labels introduced by VISIR, each for at least 10 images.  These labels are generated via generalization (from WordNet hypernyms) and abstraction (from commonsense knowledge).
As none of the baselines can produce these labels, VISIR naturally achieves a recall of 1. The precision values for the labels illustrate how VISIR addresses the problem of label incompleteness. In most cases, these labels were assessed as correct by the judges. 

\begin{table}[ht]
\centering
\caption{New labels suggested by VISIR \vspace{-0.7em}}
\label{solveIncompleteness}
\begin{tabularx}{\columnwidth}{bbs}
\hline
Label           & Label frequency   & Precision\\
\hline
individual	    &   46	            &   0.59\\
man or woman	&   44	            &   0.64\\
animal	        &   31              &   0.94\\
human           &	20              &   0.95\\
canine	        &   18              &   0.94\\
furniture	    &   12              &   0.83\\
barking animal	&   11          	&   1.00\\
\end{tabularx}
\end{table}

\section{Discussion of Limitations}
\label{discussions}

\noindent
\textbf{Training Bias in LSDA: }
The LSDA tool predicts only leaf-level object classes of ImageNet. 
The same limitation holds for most other state-of-the-art
object detectors.
Because of this incomplete tag space many important objects cannot be detected. 
For example, \textit{giraffe} is not a leaf-level object class of ImageNet. 
Since LSDA did not see 
any training images of a giraffe, it mis-labels objects 
in Figure~\ref{fig:cbir} according to its training.
This noise propagates to our model, sometimes making it impossible to find the correct labels.
\vspace*{0.2cm}

\noindent
\textbf{Incorrect Sense Mapping in ImageNet: }
LSDA trains on ImageNet images. 
Hence improper word sense mappings in ImageNet propagate 
to incorrect labels from LSDA.  
For example,
ImageNet contains similar images for two separate synsets 
\textit{Sunglass (a convex lens used to start a fire)} and 
\textit{Sunglasses (shades, dark glasses)}.
Naturally, these two synsets have completely different 
WordNet hypernyms which VISIR uses, hence introducing noise. 
The direct hypernym of \textit{Sunglass} is \textit{lens}, 
while that of \textit{Sunglasses} is \textit{glasses}.
\vspace*{0.2cm}

\noindent
\textbf{Incomplete Spatial Co-location data: }
Spatial co-location patterns mined from text contain noise 
due to linguistic variations in the form of proverbs. 
For example, 
the commonsense knowledge base WebChild~\cite{tandon2014webchild} 
assigns significant confidence to the spatial co-location of \textit{elephant} and \textit{room} (most likely from the idiom
``the elephant in the room'').
To counter such linguistic bias, we have mined spatial co-location information 
from the manually annotated ground truth of ILSVRC~\cite{ILSVRC15}.  
Unfortunately, annotations are available for only 200 object classes, leaving us with
only a small fraction of
annotated visual-label pairs.
If more cues of this kind
were available, we would have been able to establish stronger contextual coherence.
\vspace*{0.2cm}

\noindent
\textbf{Incomplete and Noisy Commonsense Knowledge: }
ConceptNet and WebChild are
quite incomplete;
so we cannot assign an abstract concept to every detected visual label. 
Also, assertions in these knowledge bases are often contradictory and noisy. 
We manage to reduce the noise by considering 
semantic relatedness with the visual labels, 
but this only alleviates part
of the problem.
\balance
\section{Conclusion}
We presented VISIR, a new method for refining and
expanding visual labels for images.
Its key strengths are cleaning out noisy labels from
predictions by object detection tools and adding informative labels
that capture generalizations and abstractions.
Our model makes this feasible by considering the visual
similarity of labels, the semantic coherence across
concepts, and various kinds of background knowledge.
The joint inference on an enriched label candidate space
is performed by means of a judiciously designed Integer Linear Program.
Our experiments show the viability of the approach,
and also demonstrate significant improvements over
two state-of-the-art object detection and tagging tools.

\section*{Acknowledgements}
This work was partly supported by the Alexander von Humboldt Foundation.

\newpage


\end{document}